\documentclass{aa}
\usepackage{txfonts}
\usepackage{graphicx}
\newcommand{\degpt}[2]{\mbox{$\rm #1\hspace{-0.25em}\stackrel{\circ}{.}
      \hspace{-1.0mm}#2$}}                          
\newcommand{\magpt}[2]{\mbox{$\rm #1\hspace{-0.25em}\stackrel{m}{.}
      \hspace{-1.0mm}#2$}}                             
\newcommand{\magpts}[2]{\mbox{$\it #1\hspace{-0.25em}\stackrel{\rm m}{.}
      \hspace{-1.0mm}#2$}}                             

\def\bsec{\hbox{$.\!\!{\arcsec}$}}

\newcommand\RA[4]{#1$^{\rm h}$#2$^{\rm m}$#3$\stackrel{\rm s}{.}$#4}

\newcommand\DEC[4]{#1$^{\circ}$#2\arcmin#3\bsec#4}
\newcommand\teff{$ {\rm T_{eff}}$}

\newcommand\logg{$\log {\rm g}$}
\newcommand\loghe{${\rm \log{\frac{n_{He}}{n_{H}}}}$}

\newcommand\ebv{$ {\rm E_{B-V}}$}
\newcommand{\Msolar}{\mbox{\,$\rm M_{\odot}$}}        
\begin{document}

\title{Observations of the Hot Horizontal-Branch Stars in the Metal-Rich
  Bulge Globular Cluster NGC~6388\thanks{Based on observations with the
  ESO Very Large Telescope 
at Paranal Observatory, Chile (proposal ID 69.D-0231(A))}
} 
\subtitle{Indications of Helium Enrichment and a Lesson in Crowded
  Field Spectroscopy} 
\author{S. Moehler\inst{1}\thanks{Present Address: European Southern
  Observatory, Karl-Schwarzschild-Str. 2, D 85748 Garching,
  Germany. {\sl e-mail: smoehler@eso.org} }
 \and A. V. Sweigart\inst{2}}

\institute{
Institut f\"{u}r Theoretische Physik und Astrophysik,
Olshausenstra\ss e 40, 24118 Kiel, Germany
\and
NASA Goddard Space Flight Center, Code 667, Greenbelt, MD 20771, USA}
\date{Received 13 April 2006/ Accepted 19 May 2006}

\abstract 
{The metal-rich bulge globular cluster NGC~6388 shows a distinct blue
horizontal-branch tail in its colour-magnitude diagram (Rich et
al. 1997) and is thus a strong case of the well-known {\it 2$^{nd}$
Parameter Problem}. In addition, its horizontal branch (HB) shows an upward
tilt toward bluer colours, which cannot be explained by canonical
evolutionary models.
Several noncanonical scenarios have been proposed to explain these
  puzzling observations. In order to test the predictions of
  these scenarios, we have obtained medium resolution spectra to determine
  the atmospheric parameters of a sample of the blue HB stars in NGC~6388.
Using the medium resolution spectra, we determine effective
  temperatures, surface gravities and helium abundances by fitting the
  observed Balmer and helium lines with appropriate theoretical
  stellar spectra. As we know the distance to the cluster, we can
  verify our results by determining masses for the stars. During the
  data reduction we took special care in subtracting the
  background, which is dominated by the overlapping spectra of cool stars.
The cool blue tail stars in our sample with effective temperatures \teff\ 
$\approx$ 10,000~K
have lower than canonical surface gravities, suggesting that these stars
are, on average, ${\approx}$0\fm4 brighter than canonical HB stars
in agreement with the observed upward slope of the HB in NGC~6388.
Moreover, the mean mass of these stars agrees well with theoretical
predictions.  In contrast, the hot blue tail stars in our sample with
\teff\ $\ge$ 12,000~K show significantly lower surface gravities than
predicted by any scenario, which can reproduce the photometric
observations.  Their masses are also too low by about a factor of 2
compared to theoretical predictions.
The physical parameters of the blue HB stars near 10,000~K
support the helium pollution scenario.
The low gravities and masses of the hot blue tail stars, however, are probably
caused by problems with the data reduction, most likely
due to remaining background light in the spectra,
which would affect the fainter hot blue tail stars much more
strongly than the brighter cool blue tail stars.  Our study of
the hot blue tail stars in NGC~6388 illustrates the obstacles
which are encountered when attempting to determine the atmospheric
parameters of hot HB stars in very crowded fields using ground-based
observations.  We discuss these obstacles and offer possible
solutions for future projects.
}

\authorrunning{Moehler \& Sweigart}
\titlerunning{Hot HB Stars in NGC~6388}
\maketitle


\section{Introduction}\label{sec:intro}
Ever since its discovery over 30 years ago (Sandage \& Wildey
\cite{sawi67}; 
van den Bergh \cite{vdbe67}), the 2$^{\rm nd}$ parameter effect has
    stood as one 
of the major unsolved challenges in the study of the Galactic globular
clusters.  While it was recognized quite early that the horizontal
branch ({\bf HB}) becomes redder on average with increasing metallicity,
many pairs of globular clusters are known with identical metallicities
but markedly different HB morphologies, e.g., M~3 versus M~13.  Thus some
parameter(s) besides metallicity (the 1$^{\rm st}$ parameter) must
affect the evolution of the HB stars in these globular clusters.
Possible 2$^{\rm nd}$ parameter candidates include the globular cluster
age, mass loss along the red-giant branch ({\bf RGB}), helium abundance $Y$,
$\alpha$-element abundance, cluster dynamics, stellar rotation, deep
mixing, etc.

Hubble Space Telescope observations by Rich et al. (\cite{riso97})
have found an unexpected population of hot HB stars in the metal-rich
globular clusters NGC~6388 and NGC~6441 ([Fe/H] $\approx-0.5$),
making these clusters the most metal-rich clusters to show the
2$^{\rm nd}$ parameter effect.  Ordinarily
metal-rich globular clusters
have only a red HB clump.  However, NGC~6388 and NGC~6441 possess
extended blue HB tails containing ${\approx} 15$\% of the total HB
population.  Quite remarkably, the HBs in both clusters slope upward
with decreasing $B - V$ with the stars at the top of the blue tail
being nearly 0\fm5 brighter in $V$ than the well-populated red HB
clump. Moreover, the RR Lyrae variables in these clusters have unusually
long periods for the cluster metallicity, leading Pritzl et al. (\cite{prsm00})
to suggest that NGC~6388 and NGC~6441 may represent a new Oosterhoff
group.  For all of these reasons the HBs of NGC~6388 and NGC~6441 are
truly exceptional.

In the next section we will discuss the implications of NGC~6388
and NGC~6441 for the 2$^{\rm nd}$ parameter effect and will review a number
of scenarios for explaining the HB morphology of these
clusters. Sect.~\ref{sec:obs} describes the observations and reduction
of the medium resolution spectra that we have obtained to test these
scenarios, while Sect.~\ref{sec:analysis} gives the atmospheric parameters
derived from these spectra.  These results
are compared with the theoretical HB tracks
in Sect.~\ref{sec:theory}. Sect.~\ref{sec:conclusions}
summarizes our conclusions. 

\section{Horizontal-Branch Morphology: Problems and Scenarios}
\label{sec:scenario}
The presence of hot
HB stars in globular clusters as metal-rich as NGC~6388 and NGC~6441 may
provide an important diagnostic for understanding the 2$^{\rm nd}$ parameter
effect for the following reason.  In intermediate-metallicity globular
clusters such as M 3 the HB spans a wide range in color that extends
both blueward and redward of the instability strip.  The location of a
star along the HB is then quite sensitive to changes in the stellar
parameters.  In fact, this is why the HB is ``horizontal''.  In
metal-rich globular clusters, however, the situation is different.
Due to their high envelope opacity, metal-rich HB stars are normally
confined to a red clump.  To move such stars blueward requires a larger
change in the stellar structure than for intermediate-metallicity
  stars.  Thus any 2$^{\rm nd}$ parameter candidate
capable of producing hot HB stars in a metal-rich globular cluster might
also have other observational consequences.  Indeed, the upward sloping
HBs in NGC~6388 and NGC~6441 suggest that the 2$^{\rm nd}$ parameter in these
globular clusters is affecting both the temperature and the luminosity of
the HB stars.

Can canonical models explain the upward sloping HBs in NGC~6388 and
NGC~6441?  In principle, one could produce hot HB stars
in these globular clusters by increasing the cluster age or by enhancing the
amount of mass loss along the RGB.  Rich et al. (\cite{riso97})
considered both of these possibilities but found neither of them
to be satisfactory because the required increase in the cluster
age is quite large and because the frequency of stellar
interactions within the cores of these clusters seems too
low to produce the additional RGB mass loss.  This conclusion
was further supported by the theoretical HB simulations of
Sweigart \& Catelan (\cite{swca98}, hereafter SC98).  They found
that the HB morphology predicted by canonical HB models
is flat in the (${M_V}$, $B-V$) plane.  Increasing
the cluster age or the RGB mass loss simply moves the models
blueward in $B-V$ without increasing their luminosity.  Thus
canonical HB models cannot account for the HB morphology of
NGC~6388 and NGC~6441.  In particular, two of the most prominent
2$^{\rm nd}$ parameter candidates -- age and RGB mass loss -- do
not work.

More recently, Raimondo et al. (\cite{raca02}) have argued that
canonical HB models can produce upward sloping HBs if the metallicity is
sufficiently high or if the mixing-length ratio $\alpha$ describing
the superadiabatic convection in the outer layers of the red
HB stars is sufficiently small.  Their solar metallicity models
with $\alpha$ = 1.0 predict a luminosity difference $\Delta V$
between the top of the blue tail and the red HB clump that exceeds
the 0\fm5 observed in NGC~6388 and NGC~6441.  However, such a
small value for the mixing-length ratio is untenable because
it would imply a $B - V$ colour for the RGB that is much redder
than observed.  The predicted $\Delta V$ for the more appropriate
mixing-length ratio $\alpha$ = 1.6 is only ${\approx}$ 0\fm2
for a solar metallicity and decreases further with
decreasing [Fe/H].  At the metallicity of NGC~6388 and
NGC~6441 the Raimondo et al. models predict little difference
in the luminosity of HB stars at the top of the blue tail
compared to those in the red clump, further confirming that
some noncanonical effect must be influencing the HB
morphology in NGC~6388 and NGC~6441.

The failure of canonical HB models to produce upward sloping HBs
has promp\-ted the study of other noncanonical solutions.  Theoretical
models show that the HB luminosity at a fixed metallicity depends
on two parameters: the helium abundance $Y$ and the core mass ${M_c}$.  This
fact lead SC98 to suggest 3 noncanonical scenarios involving increases
in either $Y$ or ${M_c}$ which might potentially produce
upward sloping HBs.

The first (``high-$Y$'') scenario assumes that the stars in
NGC 6388 and NGC 6441 formed with a high primordial helium
abundance due to a peculiar chemical enrichment history in these
clusters.  From theoretical models we know that HB tracks
at high helium abundances have very long blue loops which
deviate considerably from the zero-age HB ({\bf ZAHB}).  The HB
simulations of SC98 show that such high-$Y$ tracks can indeed
produce upward sloping HBs as seen in NGC~6388 and NGC~6441
provided $Y$ is very large 
(\raisebox{-0.5ex}{$\stackrel{>}{\scriptstyle \sim}$}\,0.4).  
However, this
scenario
predicts too large a value for the number ratio $R$ of
HB stars to RGB stars brighter than the HB (Layden et al. \cite{lari99})
as well as too bright a luminosity for the RGB bump (Raimondo et
al. \cite{raca02}).  Thus a high
primordial helium abundance in NGC~6388 and NGC~6441 can be ruled out.

The second (``rotation'') scenario is based on the fact that internal
rotation within an RGB star can delay the helium flash, thereby
leading to a larger core mass and greater mass loss near the tip of
the RGB.  This increase in the core mass ${M_c}$ together with
the corresponding decrease in the total mass $M$ will shift a
star's HB location towards higher effective temperatures and
luminosities.  HB simulations show that this scenario can also produce
upward sloping HBs similar to those observed in NGC~6388 and NGC~6441.
However, it is difficult to understand how the blue HB stars in
NGC~6388 and NGC~6441 could have the high rotation rates required by
this scenario. Preliminary results from high resolution spectra of
three cool blue HB stars (\teff\ $\lesssim$ 10,000~K) in NGC~6388 do
not show evidence for rotation velocities of $v \sin i \gtrsim 10$ km
sec$^{-1}$ (Moehler \& Sweigart \cite{mosw06}).  Thus the rotation
scenario seems unlikely.

The third (``helium-mixing'') scenario is motivated by the large
star-to-star abundance variations which are found among the red-giant
stars within individual globular clusters and which are sometimes attributed
to the mixing of nuclearly processed material from the
vicinity of the hydrogen shell out to the stellar surface (Kraft
\cite{kraf94}).  The 
observed enhancements in Al are particularly important because they
would require the mixing to penetrate deeply into the hydrogen
shell (Cavallo et al. \cite{casw98}).  Such mixing would dredge up fresh helium
together with Al, thereby increasing the envelope helium abundance
and leading to a brighter RGB tip luminosity and hence greater
mass loss.  Thus
a helium-mixed star would arrive on the HB with both a higher
envelope helium abundance and a lower mass and would therefore
be both bluer and brighter than its canonical counterpart - just
what is needed to produce an upward sloping HB.  Indeed, the
HB simulations of SC98 confirm that helium mixing can produce
HB morphologies similar to those in NGC~6388 and NGC~6441.  However,
the existence of helium mixing on the RGB can be questioned on several
grounds.  The O-Na and Mg-Al anticorrelations observed in turnoff
stars of NGC~6752 by Gratton et al. (\cite{grbo01}) indicate that
the Al enhancements are more likely due to primordial pollution from an earlier
generation of stars
than to deep mixing on the RGB.  It
is also questionable whether the mixing currents could overcome the large
gradient in the mean molecular weight within the hydrogen shell
of a RGB star.  Thus
helium mixing also seems unlikely.

These difficulties have prompted a number of additional suggestions
for explaining the HB morphologies of NGC~6388 and NGC~6441.  One of
the earliest suggestions, offered by Piotto et al. (\cite{piso97}),
was a spread in metallicity.  In this case the blue HB stars would be
metal-poor compared to the stars in the red HB clump.  Because the HB
becomes brighter in $V$ with decreasing metallicity, the blue HB stars
would also be brighter.  Thus a metallicity spread might also produce
an upward sloping HB.  This possibility was studied by Sweigart
(\cite{swei02}), who showed that upward sloping HBs similar to those
in NGC~6388 and NGC~6441 would require the stars at the top of the
blue HB tail to be approximately 2 dex more metal-poor than the stars
in the red HB clump.  However, Raimondo et al. (\cite{raca02}) have
noted that the progenitors of the blue HB stars should appear as a
population of metal-poor giants lying well to the blue of the
metal-rich RGB.  Since such metal-poor giants are not seen in the
colour-magnitude diagrams of NGC~6388 and NGC~6441, Raimondo et
al. (\cite{raca02}) conclude that any metallicity spread must be
small. Again preliminary results of the high resolution spectra
mentioned above do not show any evidence for a significant metal
deficiency in these blue HB stars compared to the overall metallicity
of NGC~6388 (Moehler \& Sweigart \cite{mosw06}).

Piotto et al. (\cite{piso97}) also suggested that NGC~6388 and NGC~6441 might
contain two stellar populations with different ages.  This
possibility has been further explored by Ree et al. (\cite{reyo02}).  Their
population models for NGC~6388 and NGC~6441 are able to produce blue HB
stars provided these stars are older by 1.2 Gyr and metal-poor
by 0.15 dex compared to the stars in the red HB clump.  Such
a small difference in metallicity between the blue and red HB
stars avoids the problem with the missing metal-poor giants
discussed above.  However, while
such models might produce bimodal HBs, they do not produce an
upward sloping HB.  As shown by SC98, differences in age
merely move an HB star horizontally along the HB, and, as shown
by Sweigart (\cite{swei02}), a metallicity difference of 0.15 dex is
too small to produce a significant HB slope.  Thus the Ree
et al. (\cite{reyo02}) models do not account for a key property
of the HBs in NGC~6388 and NGC~6441.  Ree et al. (\cite{reyo02}) also
suggest that the long RR Lyrae periods might be explained
if these stars are highly evolved from the blue
HB.  While such stars would be brighter and hence have longer periods,
they would also evolve rapidly across the instability strip
on their way to the asymptotic-giant branch
({\bf AGB}).  Explaining the observed number of RR Lyrae stars
under such a scenario would therefore be very difficult (Pritzl et
al. \cite{prsm02}). 

\begin{table*}
\caption[]{Coordinates and photometric data for our
  target stars (from Piotto et al. \cite{piso97})\label{ngc6388phot}} 
\begin{tabular}{rllrrr}
\hline
\hline
 number & \multicolumn{1}{c}{$\alpha_{2000}$} & 
\multicolumn{1}{c}{$\delta_{2000}$} & 
\multicolumn{1}{c}{$V$} & \multicolumn{1}{c}{$B-V$} & 
\multicolumn{1}{c}{setup} \\
\hline
 1283 & \RA{17}{36}{23}{899} & \DEC{-44}{44}{58}{45} &
 \magpt{18}{690}$\pm$\magpt{0}{032} & \magpt{+0}{230}$\pm$\magpt{0}{041}
 & 1\\
 1775 & \RA{17}{36}{14}{921} & \DEC{-44}{43}{08}{87} &
 \magpt{18}{811}$\pm$\magpt{0}{040} & \magpt{+0}{292}$\pm$\magpt{0}{047}
 & 1\\
 2612 & \RA{17}{36}{22}{560} & \DEC{-44}{44}{01}{93} &
 \magpt{19}{622}$\pm$\magpt{0}{059} & \magpt{+0}{279}$\pm$\magpt{0}{068}
 & 1\\
 4113 & \RA{17}{36}{22}{884} & \DEC{-44}{44}{24}{63} &
 \magpt{16}{776}$\pm$\magpt{0}{026} & \magpt{+0}{545}$\pm$\magpt{0}{037}
 & 1\\
 6239 & \RA{17}{36}{19}{813} & \DEC{-44}{43}{05}{88} &
 \magpt{19}{147}$\pm$\magpt{0}{036} & \magpt{+0}{244}$\pm$\magpt{0}{041}
 & 1\\
 7104 & \RA{17}{36}{17}{879} & \DEC{-44}{45}{05}{02} &
 \magpt{19}{310}$\pm$\magpt{0}{026} & \magpt{+0}{230}$\pm$\magpt{0}{037}
 & 1\\
 7788 & \RA{17}{36}{21}{842} & \DEC{-44}{42}{55}{70} &
 \magpt{17}{410}$\pm$\magpt{0}{023} & \magpt{+0}{360}$\pm$\magpt{0}{026}
 & 1\\
  396 & \RA{17}{36}{21}{160} & \DEC{-44}{44}{24}{42} &
 \magpt{17}{383}$\pm$\magpt{0}{025} & \magpt{+0}{303}$\pm$\magpt{0}{029}
 & 2\\
 1233 & \RA{17}{36}{24}{565} & \DEC{-44}{43}{15}{21} &
 \magpt{16}{867}$\pm$\magpt{0}{029} & \magpt{+0}{428}$\pm$\magpt{0}{038}
 & 2\\
 2483 & \RA{17}{36}{21}{915} & \DEC{-44}{44}{08}{02} &
 \magpt{16}{927}$\pm$\magpt{0}{024} & \magpt{+0}{467}$\pm$\magpt{0}{028}
 & 2\\
 2897 & \RA{17}{36}{23}{334} & \DEC{-44}{43}{56}{46} &
 \magpt{18}{835}$\pm$\magpt{0}{029} & \magpt{+0}{264}$\pm$\magpt{0}{038}
 & 2\\
 5235 & \RA{17}{36}{21}{301} & \DEC{-44}{45}{14}{05} &
 \magpt{17}{113}$\pm$\magpt{0}{019} & \magpt{+0}{341}$\pm$\magpt{0}{024}
 & 2\\
 7567 & \RA{17}{36}{16}{614} & \DEC{-44}{44}{59}{74} &
 \magpt{18}{882}$\pm$\magpt{0}{145} & \magpt{+0}{343}$\pm$\magpt{0}{158}
 & 2\\
\hline
\end{tabular}
\end{table*}

\begin{table*}
\caption[]{Observation parameters for the spectroscopic
  data. The FORS2 value of the seeing has been measured on the
  acquisition image taken immediately before the spectroscopic
  exposures. Between the first and second observation of setup 2 no
  acquisition image was taken. \label{tab:obs}}
\begin{tabular}{cccccccrr}
\hline
\hline
setup & date & start of  & exposure & \multicolumn{2}{c}{seeing} & airmass &
\multicolumn{2}{c}{moon} \\
      &      & exposure  & time     & DIMM & FORS2 &        & illumination &
distance \\
\hline
1 & 2002-07-10 & 03:36:27.049 UT & 3320 sec & 0\farcs91 & 0\farcs75 &
1.095 & 0.1\% & \degpt{153}{7} \\
 & 2002-08-02 & 03:37:04.339 UT & 3238 sec & 1\farcs78 & 1\farcs25 & 
1.245 & 42.5\% & \degpt{136}{6}\\
2 & 2002-08-02 & 01:34:19.337 UT & 3238 sec & 1\farcs70 & 1\farcs00 & 
1.075 & 43.5\% & \degpt{135}{5} \\
  &            & 02:31:37.199 UT& 3238 sec & 1\farcs75 & --- & 
1.121 & 43.0\% & \degpt{136}{0} \\
\hline
\end{tabular}
\end{table*}

\begin{figure}[h]
\includegraphics[width=\columnwidth]{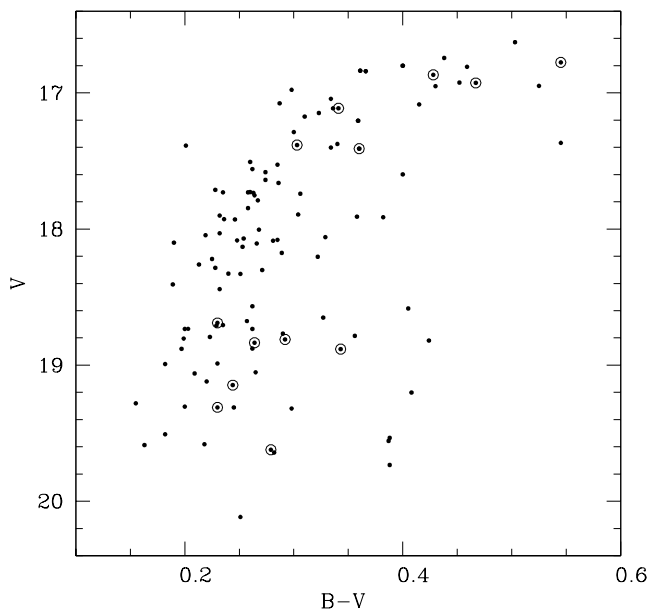}
\caption[]{The colour-magnitude diagram of the hot horizontal-branch stars in
  NGC~6388 (Piotto et al. \cite{piso97})
with our spectroscopic targets marked by open circles.\label{fig:cmd}}
\end{figure}
A more promising possibility is based on the suggestion by
D'Antona \& Caloi (\cite{daca04}) that the stars in globular clusters
with blue HB tails 
are born in two events: a first generation of helium-normal stars
and a second generation of helium-rich stars which form from
the ejecta of the first generation.  Various
candidates have been proposed for producing this helium-rich
ejecta, including AGB stars, massive main-sequence stars,
type II supernovae, and even stars external to the
globular cluster (Bekki \& Norris \cite{beno06} and
references therein).  Support for such helium pollution
comes from the recent discoveries of a double main sequence
in $\omega$ Cen (Anderson \cite{ande97};
Norris \cite{norr04}; Piotto et al. \cite{pivi05}) and
a blueward extension of the main sequence in NGC~2808
(D'Antona et al. \cite{dabe05}).  These globular clusters
apparently contain a significant population of
helium-rich ($Y \approx 0.4$) stars.
Since a helium-rich star has a lower turnoff mass
at a given age, it will be bluer on the HB than a helium-normal
star.  It will also be brighter due to the increased energy
output of the hydrogen-burning shell.  Thus the spread in 
the helium abundance predicted by this helium pollution scenario will
lead to a spread in color along the HB, with the red clump
stars corresponding to the helium-normal, first generation
stars and the blue tail stars being progressively more helium-rich
as the effective temperature increases.  An upward sloping HB
is a natural consequence of this spread in helium.  The fact
that the HB slope is more prominent in NGC~6388 and NGC~6441 than
in other blue tail globular clusters may be simply due to their higher
metallicity which requires a larger increase in helium in order
to force a star blueward of the red clump.  We emphasize
that this scenario differs from the high-$Y$ scenario, mentioned
above, in which all of the stars are helium-rich and from the
helium-mixing scenario in which the spread in helium arises
from deep mixing on the RGB.

All of the scenarios which can produce an upward sloping HB
predict that the gravities of the blue HB stars should be lower than the
gravities of canonical blue HB stars.  In 1998 we observed some of the
brighter objects in both clusters ($B<$18; four stars in NGC~6388, three
in NGC~6441) to test this prediction. Unfortunately the results only
added to the confusion as most of the stars had even {\it higher gravities}
than predicted by canonical evolution (Moehler et al. \cite{mosw99}). In
2002 we obtained additional medium resolution
spectra of about a dozen blue HB stars along the blue tail
of NGC~6388 to determine
their effective temperatures and surface gravities. In
the following sections we will describe our analysis of these
spectra and will compare the derived effective temperatures
and gravities with the predictions of the
helium pollution scenario.

\begin{figure*}
\begin{minipage}{\columnwidth}{
\caption[]{Finding charts for all target stars from a FORS2 image
  obtained in white light with a seeing of 0\farcs6.
\label{fig:find}}}\end{minipage}
\end{figure*}

\section{Observations and Data Reduction}\label{sec:obs}
\subsection{Target Selection}
 Our spectroscopic targets were selected from the catalog of
Piotto et al. (\cite{piso97}; Fig.\ref{fig:cmd}), and include six stars in
the roughly horizontal part of the horizontal branch (\magpt{16}{6} $< V
<$ \magpt{17}{6}) and seven stars
at the hot end of the blue tail (\magpt{18}{6} $< V <$
\magpt{19}{8}). We tried to select the most isolated stars from the
original WFPC2 images, which were obtained close to the core of the
globular cluster.
The coordinates and photometry for our targets
are given in Table~\ref{ngc6388phot}. Fig.~\ref{fig:find} shows
  finding charts for the stars.

\subsection{Spectroscopy}
We obtained medium-resolution spectra ($R\approx$1200)  at
the VLT-UT4 (Yepun) with FORS2 in service mode
(see Table~\ref{tab:obs} for details).  We used the multi-object
spectroscopy (MOS) mode of FORS2 (slit length 20\arcsec respectively
22\arcsec) with the
standard collimator (0\bsec2/pixel), a slit width of 0\farcs6 and
grism B600. The slitlets were positioned to cover at least the
wavelength range of 3700~\AA\
to 5200~\AA. As FORS2 is equipped with an atmospheric dispersion
corrector, MOS observations at higher airmass are not a problem. 
Due to a mistake during the Phase 2 preparations the mask for the second
observation of setup 1 was prepared with 1\farcs2 wide slitlets. In
combination with the mediocre seeing these wide slits made the data
useless due to the high crowding evident in Fig.~\ref{fig:find}.
Therefore we have only one spectrum for each star in setup 
1. 

FORS2 has two 2k$\times$4k thinned MIT CCD detectors with
anti-reflection coating and a pixel size of (15$\mu$m)$^2$ with a gain
of 0.7 e$^-$/count and a read-out noise of 2.7 e$^-$. The CCDs are
read out with binning 2$\times$2, resulting in a scale of
0\farcs25/pixel respectively 1.5\AA/pixel. Due to the limited
field-of-view of the WFPC2 photometry, from which we selected our
targets, we used only the master CCD for our observations.

For each night screen flat fields with two different illumination
patterns and CdHeHg wavelength calibration spectra were observed. As
part of the standard calibration we were also provided with masterbias
frames for our data. The masterbias showed no evidence for hot pixels
and was smoothed with a 30$\times$30 box filter to keep any possible
large scale variations while erasing noise. The flat fields were averaged
for each night and bias-corrected by subtracting the smoothed
masterbias of that night. From the flat fields we determined the
limits of the slitlets in spatial direction. Each slitlet was
extracted and from there on treated like a long-slit spectrum. The
flat fields were normalized with 4$^{\rm th}$ to 7$^{\rm th}$ order
polynomials. The dispersion relation was obtained from the wavelength
calibration frames by fitting 4$^{\rm rd}$ to 6$^{\rm th}$ order
polynomials to the line positions along the dispersion axis. We used
15 to 19 unblended lines between 3600~\AA\ and 5900~\AA\ and achieved
a typical r.m.s. error of 0.05~\AA\ to 0.08~\AA\ per CCD row.

Due to exposure times of more than 50 minutes the scientific observations
contained a large number of cosmic ray hits. Those were corrected with
the algorithm described in Pych (\cite{pych04}) with the default parameters.
To ensure that this procedure did not introduce any artefacts,
 we also reduced the uncorrected frames for comparison. We did not find
 any problems with the cosmic ray rejection routine. The
slitlets with the stellar spectra were extracted in the same way as
the flat field and wavelength calibration slitlets. The smoothed
masterbias was subtracted, and the spectra were divided by the
corresponding normalized flat fields, before they were rebinned
2-dimensionally to constant wavelength steps. We then corrected the
curvature of the spectra along the spatial axis using a routine by
O. Stahl (priv. comm.), which traces a spectrum in a predefined spatial
region and determines and corrects the curvature derived that way for
the complete slitlet. If possible (i.e. if the target was sufficiently
isolated and/or sufficiently bright) we used the target spectrum for the
curve tracing, otherwise the brightest spectrum in the slitlet.

\subsection{Sky Subtraction}\label{sec:skysub}
\begin{figure}[h]
\includegraphics[height=\columnwidth,angle=270]{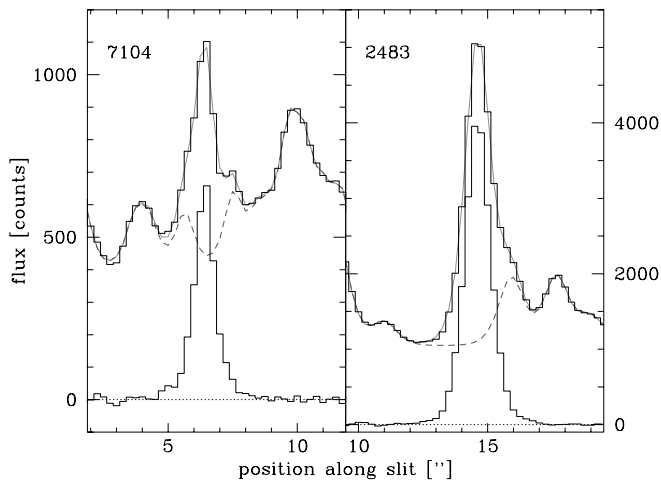}
\caption[]{The spatial light distribution along two slitlets (left: star
  7104, setup 1; right: star 2483, setup 2). The upper 
histogram is the observed light distribution, the grey continuous line
gives the fit of all sources, the grey dashed line marks the subtracted
sky background (= fit of all sources except target). The lower histogram is
the spatial profile of the sky subtracted image.\label{fig:prof_exam}}
\end{figure}

The MOS observations were obtained in very crowded fields, and due to
the fact that several stars were observed simultaneously we could not
orient the individual slitlets in a way to avoid nearby stars. Therefore
most slitlets contain spectra of several stars, in many cases
overlapping so strongly that it is impossible to directly extract the
spectrum of our intended target (see Fig.~\ref{fig:prof_exam}). In order
to account for this overlap, we proceeded as follows:
\begin{enumerate}
\item we corrected the curvature of the FORS2 spectra 
\item we averaged the wavelength-calibrated
two-dimensional spectra along their dispersion axis between 3500~\AA\
and 5200~\AA\ (roughly the range which is later used for fitting the
line profiles), thereby producing a
one-dimensional spatial profile along the slitlet
(cf. Fig.~\ref{fig:prof_exam}, upper histogram) 
\item the one dimensional spatial distribution of light was fitted with
  a combination of Moffat functions, i.e.
\[ I_{(x)} = bck + \sum_{j = 1}^{n} I_{j(x)} \]
with
\[I_{j(x)} = a_j \left (1+ \frac{4 (x-b_j)^2}{c^2} \right )^{-d}  \]
For each profile the parameters $a_j$ (amplitude) and $b_j$ (position) were
fitted individually, whereas the parameters $c$ and $d$, which determine
the profile width and shape (and should depend only on the seeing and
instrumental broadening), had to be the same for all objects within
one slitlet. We used up to 13 individual profiles to fit the full
spatial light distribution along one slitlet (see
Fig.~\ref{fig:prof_exam}, light solid line).  
\item After achieving a good fit
to the observed spatial profiles we kept the parameters $b_j$, $c$, and
$d$ fixed and used these profiles to fit the spatial profile now at
every wavelength step. The amplitudes $a_j$ and the spatial constant
$bck$ (for the true sky background) were allowed to vary with
wavelength, in order to describe the various spectra.
\item The sum of all profiles {\it except the target profile} was then
 used as background (cf. Fig.~\ref{fig:prof_exam}, dashed line) and
 subtracted from the wavelength calibrated two-dimensional image. The
 resulting image -- containing only the target spectrum -- was again
 averaged over the same wavelength range as above to verify the quality
 of the sky correction (see Fig.~\ref{fig:prof_exam}, lower histogram).
\end{enumerate}

 The sky-subtracted spectra were extracted using Horne's
(\cite{horn86}) algorithm as implemented in MIDAS. As is well-known,
 optimum extraction procedures fail for data with good S/N. We
 therefore also extracted the target spectra by simple averaging and
 compared the results from the two procedures in order to avoid
 artefacts introduced by optimum extraction. Finally the
spectra were corrected for atmospheric extinction using the extinction
coefficients for La Silla (T\"ug \cite{tueg77}) as implemented in
MIDAS, because they provide the closest approximation to Paranal
conditions, for which no spectroscopic extinction coefficients are
available.

\begin{figure}[h]
\hspace*{-3ex}
\includegraphics[width=0.5\textwidth,angle=0]{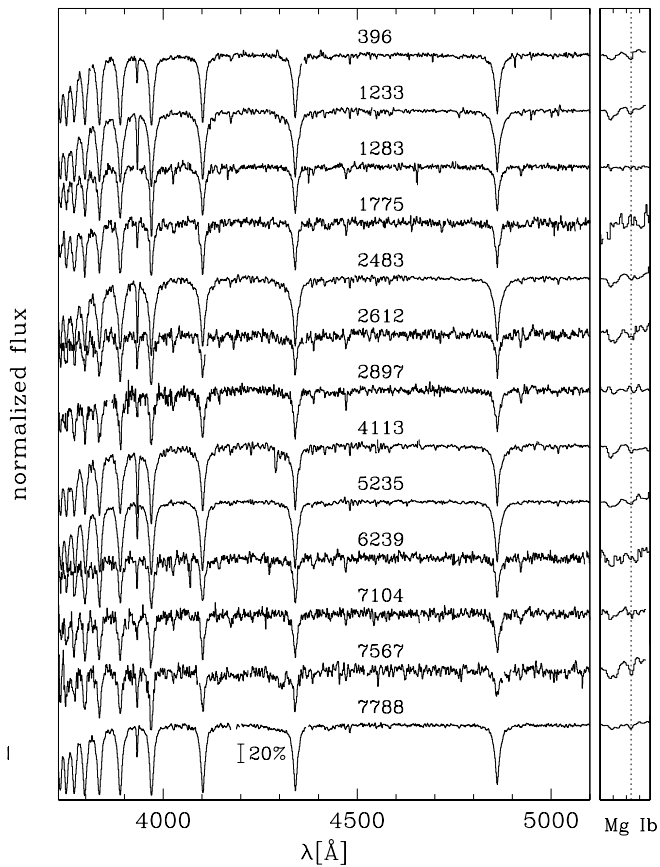}
\caption[]{Normalized spectra for all target stars. Spectral regions
  affected by cosmics are left out. On the right we show the region of
  the \ion{Mg}{1}b triplet, which is quite prominent in cool stars and
  may indicate contamination of our spectra by such stars. \label{fig:spectra}}
\end{figure}

For a relative flux calibration we used response curves derived from
spectra of LTT~6248 with the data of Hamuy et
al. (\cite{hawa92}). The response curves were fit by splines and
averaged for the two nights.  
The individual target spectra were
corrected for Doppler shifts determined from Balmer and helium
absorption lines and co-added if both available spectra were of
comparable quality. Finally all spectra were corrected for an
interstellar extinction of \magpt{0}{37}. In order to allow easy
comparison of the spectra, we normalized them using the model spectra
provided by the fit procedure (see Sect.~\ref{sec:analysis}).
These normalized spectra are shown in Fig.~\ref{fig:spectra}.

Star 7567 shows a prominent G-band and a clear \ion{Mg}{1}b absorption,
which in combination with its very red colour and high photometric
error, suggest that this star has a cool companion (either a physical
one or an unresolvable blend). It is therefore excluded from the further
analysis and discussion.

\section{Analysis}\label{sec:analysis}

\subsection{Atmospheric Parameters}

 For the analysis of the FORS2 spectra we used two sets of ATLAS9
model atmospheres (Kurucz \cite{kuru93}): For stars with $V <
$ \magpt{17}{5}, which are presumably cooler than 11,000~K and
therefore not affected by diffusion, we used a metallicity of [M/H] =
$-0.5$ and solar helium abundance. For the fainter stars with $V >
$ \magpt{18}{5}, which are presumably hotter than 12,000~K and
therefore affected by diffusion, we used a super-solar metallicity of
[M/H] = $+0.5$ and helium abundances ranging from solar to 1/100
solar. These models should account in a rough way for the effects of
radiative levitation of heavy elements and gravitational settling of
helium (see Moehler et al. \cite{mosw00} for details). From these
model atmospheres we calculated spectra with Lemke's
version\footnote{For a description see
http://a400.sternwarte.uni-erlangen.de/$\sim$ai26/linfit/linfor.html}
of the LINFOR program (developed originally by Holweger, Steffen, and
Steenbock at Kiel University). To establish the best fit, we used the
routines developed by Bergeron et al.\ (\cite{besa92}) and Saffer et
al.\ (\cite{sabe94}), as modified by Napiwotzki et
al. (\cite{nagr99}), which employ a $\chi^2$ test. The $\sigma$
necessary for the calculation of $\chi^2$ is estimated from the noise
in the continuum regions of the spectra.  The fit program normalizes
model spectra {\em and} observed spectra using the same points for the
continuum definition.  We used the Balmer lines H$_\beta$ to H$_{12}$
(excluding H$_\epsilon$ to avoid the \ion{Ca}{ii}~H line) for the
fit. For the hot stars we included also the \ion{He}{i} lines
$\lambda\lambda$ 4026~\AA, 4388~\AA, 4471~\AA, 4921~\AA\, to determine
helium abundances, whereas the helium abundance was kept fixed at the
solar value for the cool stars. The results are given in
Table~\ref{tab:par} and plotted in Fig.~\ref{fig:tg}.

Recent tests have shown, however, that these fit routines underestimate
the {\em formal} errors by at least a factor of 2 (Napiwotzki 2005,
priv. comm.). In addition, the
errors provided by the fit routine do not include possible systematic
errors due to, e.g., flat field inaccuracies or imperfect sky
subtraction. As we cannot estimate these systematic errors, we
multiplied the formal errors by 2 to provide at least a lower limit to
the errors.

\subsection{Reddening}
Using the effective temperatures derived this way, we also estimated
reddenings by comparing the predicted colour from ATLAS9 model colours
(for the same metallicities as the model spectra used in the fits) to
the observed colours. Table~\ref{tab:par} shows an obvious and very
suspicious trend, namely, that the reddening increases towards higher
effective temperatures: The cool stars show an average reddening of
\magpt{0}{35}$\pm$\magpt{0}{05}, which is consistent with the literature
value for NGC~6388. However, the fit results for stars below 9000~K are
uncertain, as such stars show more and more metal lines, which are not
included in the model spectra and possibly not evident in the observed
spectra due to the low resolution of the data (although some lines can
be seen in the spectra of the reddest stars 4113, 2483, and 1233 in
Fig.~\ref{fig:spectra}). Restricting the averaging to the stars around
10,000~K yields a mean reddening of \magpt{0}{39}. The hot stars,
however, show an average reddening of \magpt{0}{45}$\pm$\magpt{0}{04},
which is considerably larger than the accepted value.  While Raimondo et
al. (\cite{raca02}) report on possible differential reddening of up to
\magpt{0}{1}, the correlation between reddening and temperature looks
suspicious and might indicate an overestimate of the effective
  temperatures in the hot stars. As discussed below such an
  overestimate may be due to a partial filling of the line cores by
  residual light from cool stars (see Sect.~\ref{blue_tail_hot}).

\begin{table}
\caption{The atmospheric parameters and estimated reddenings from the
  spectroscopic analyses of our target stars. The spectroscopic
 results for stars below 9000~K (italized in table)
  are considered uncertain. For all stars below 11,000~K we assumed
  a solar atmospheric helium abundance, as these stars should not be
  affected by diffusion (see text for details).\label{tab:par}}
\begin{tabular}{rrccc}
\hline
\hline
star & \multicolumn{1}{c}{\teff} & \logg & \loghe & ${\rm E_{B-V}^{spec}}$ \\
     & \multicolumn{1}{c}{[K]}   & [cm s$^{-2}$]  &      & \\
\hline
 396  &  10200$\pm$\ \ 300 &  3.52$\pm$0.12 &  &  \magpt{0}{360}\\
1233 &   {\it 7640$\pm$\ \ 110} &  {\it 2.85$\pm$0.12} &  & \magpts{0}{287}\\
 1283 & 17200$\pm$\ \ 760 & 4.28$\pm$0.12 & $-$1.71$\pm$0.22 & \magpt{0}{408}\\
 1775 & 16100$\pm$\ \ 880 & 4.04$\pm$0.18 & $-$1.68$\pm$0.32 & \magpt{0}{464}\\
 2483 & {\it 7780$\pm$\ \ 110} & {\it 2.97$\pm$0.12} & & \magpts{0}{345}\\
 2612 & 24800$\pm$3300 &  4.68$\pm$0.38 & $-$1.66$\pm$0.20 & \magpt{0}{511}\\ 
 2897 & 18700$\pm$1400 &  4.59$\pm$0.20 & $-$1.48$\pm$0.32 & \magpt{0}{451}\\ 
 4113 & {\it 7200$\pm$\ \ 140} & {\it 2.72$\pm$0.22} &  & \magpts{0}{323}\\
 5235 &   9960$\pm$\ \ 320 &  3.61$\pm$0.16 &  & \magpt{0}{385}\\
 6239 &  19300$\pm$1900 &  4.33$\pm$0.26 & $-$1.55$\pm$0.26 & \magpt{0}{443}\\
 7104 &  19500$\pm$2000 &  4.43$\pm$0.24 & $-$2.06$\pm$0.36 & \magpt{0}{429}\\
 7788 &  10200$\pm$\ \ 280 &  3.51$\pm$0.14 &  & \magpt{0}{360}\\
\hline
\end{tabular}
\end{table}

At least a part of the correlation might
be due to the different effective temperatures of the stars. As
discussed by Grebel \& Roberts (\cite{grro95}) and others, the reddening
effect of interstellar matter depends on the spectral type of a star. On
average the reddening increases for bluer spectral types. Thus one would
expect blue HB stars to have a higher reddening than the red giants in
the same globular cluster. However, using their Fig.~9, we estimate that
the reddening difference between red giants with [M/H] = $-0.5$ and blue
tail stars with an atmospheric metallicity of [M/H] = $+0.5$ is at most
about \magpt{0}{04} for a reddening of \ebv\ = \magpt{0}{37}
and the difference between cool blue HB stars and
hot blue tail stars is below \magpt{0}{01}. This effect could
therefore explain the slightly higher reddening of the cooler blue HB
stars compared to the reddening of the cluster in general. 

However, from the colour-magnitude diagram in Fig.~\ref{fig:cmd}, it is
obvious that our ``hot'' targets have a higher reddening than the
``cool'' ones, as they lie all on the red side of the blue tail and in
some cases even overlap with the cooler and brighter targets in $B-V$.

\section{Comparison with Theoretical Tracks}\label{sec:theory}

\begin{figure}[h]
\includegraphics[height=\columnwidth,angle=270]{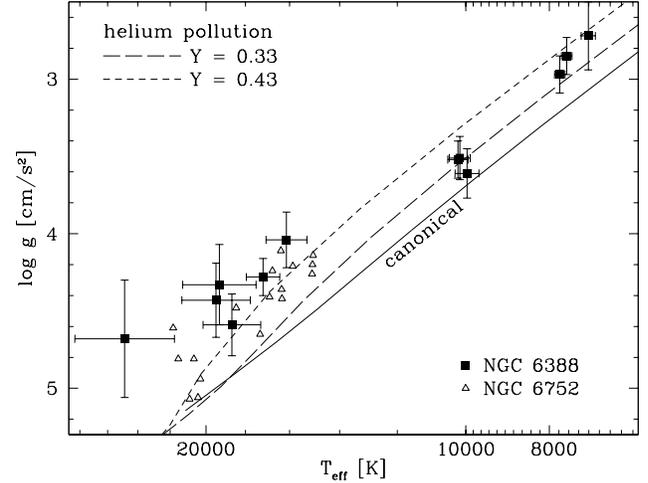}
\caption[]{Effective temperatures and surface gravities for our target
  stars as derived from line profile fits. For comparison we show the
  zero-age HB (ZAHB) for both a canonical helium abundance ($Y$ =
  0.23, solid line) and two helium-rich compositions ($Y$ = 0.33, long
  dashed line; $Y$ = 0.43, short dashed line). For comparison we
  also show the results for the metal-poor globular cluster NGC~6752
  from Moehler et al. (\cite{mosw00}).\label{fig:tg}}
\end{figure}

We next compare the derived atmospheric parameters given
in Table~\ref{tab:par} with the theoretical HB tracks for both
canonical and helium-rich compositions.  Our goals are to
determine if the blue tail stars provide any support for the helium
enhancement predicted by the helium pollution scenario and, more
specifically, to see if the stars near the top of the blue tail
have the lower than canonical gravities inferred from the upward
slope of the HB in NGC~6388.  In Fig.~\ref{fig:tg} we plot the effective
temperatures and surface gravities of our target stars together
with the ZAHBs for a canonical helium abundance
($Y$ = 0.23) and two helium-rich compositions ($Y$ = 0.33, 0.43),
as might arise from helium pollution.  According to the helium
pollution scenario, the helium abundance should increase along
the HB from its canonical value in the reddest HB stars to a
helium-rich value along the blue tail.  Thus
the blue tail stars should show the maximum offset from the
canonical surface gravities.

The stars in Fig.~\ref{fig:tg} fall into
three groups: three stars cooler than 9000~K, three stars near 10,000~K
and six stars hotter than 12,000~K.  As discussed in the previous section,
the coolest stars with \teff\
below 9000~K have less reliable parameters and thus will not be
considered further.

\subsection{Cool Blue Tail Stars Near 10,000~K}

\begin{figure}[h]
\includegraphics[height=\columnwidth,angle=270]{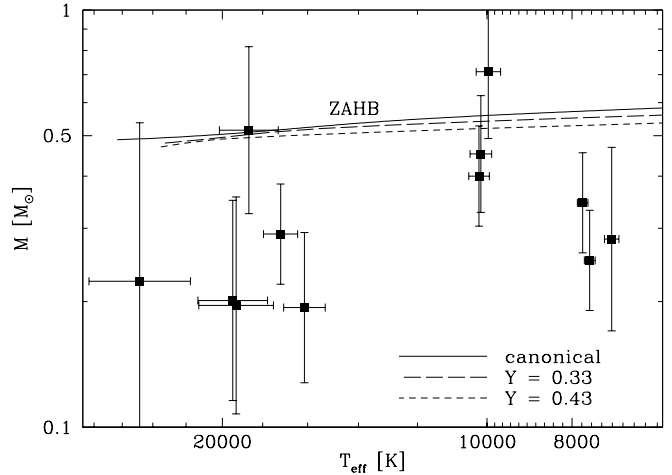}
\caption[]{Effective temperatures and masses for our target
stars.  For comparison we show both the canonical ZAHB
($Y$ = 0.23, solid line) and
two helium-rich ZAHBs ($Y$ = 0.33, long dashed line; $Y$ = 0.43,
short dashed line).\label{fig:tm}}
\end{figure}

The moderately cool stars near 10,000~K lie above the canonical ZAHB
in Fig.~\ref{fig:tg}.  At first glance this might be interpreted as
evidence for a helium enhancement in these stars of perhaps as much as
$Y$ $\approx$ 0.33.  However, before drawing such a conclusion, we need
to determine if the masses of these stars, as obtained from their
atmospheric parameters, are consistent with the theoretical HB masses.
Previous experience with the analysis of hot stars in globular
clusters has shown that low surface gravities are often associated
with impossibly small HB masses, thus indicating that the surface
gravities are not reliable.  To carry out this consistency check, we
have estimated masses for all of our target stars using the approach
outlined by Moehler et al. (\cite{mosw00}).  Fig.~\ref{fig:tm}
compares these masses with the theoretical ZAHB masses for both
canonical and helium-rich compositions.  We note that the ZAHB masses
in Fig.~\ref{fig:tm} depend only weakly on the helium abundance.  The
mean mass of the cool blue tail stars is 0.52~\Msolar\ in good
agreement with the theoretical masses.

\begin{figure}[h]
\includegraphics[width=\columnwidth,angle=0]{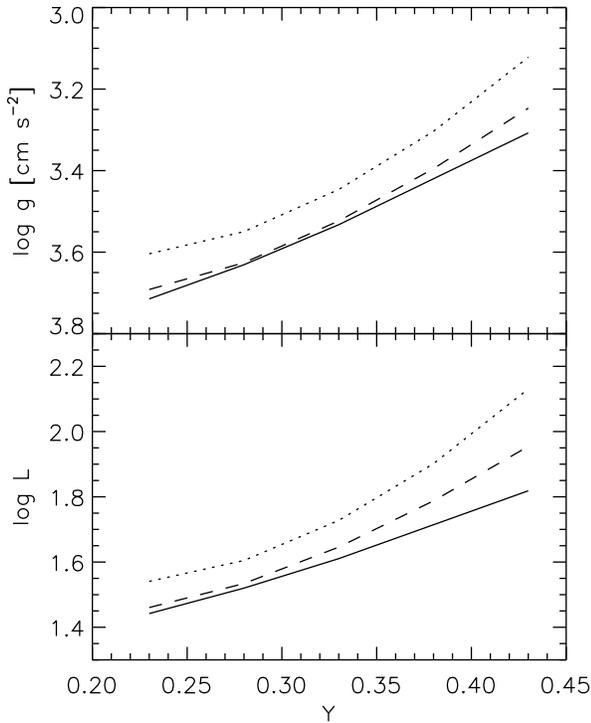}
\caption[]{Predicted surface gravity and luminosity of HB stars
at \teff\ = 10,120 K as a function of
the helium abundance $Y$.  The solid curves refer
to the ZAHB, the dashed curves to the midpoint of the HB phase,
and the dotted curves to the point 90\% through the HB phase.
\label{fig:L_g_Y}} 
\end{figure}

Given this consistency check, we can now examine how the predicted
surface gravity and luminosity of the cool blue tail stars depend
on the helium abundance.  In order to reduce the uncertainty
in our results, we will use the average effective temperature and
surface gravity of these stars instead of the individual measurements,
i.e., $<$\teff $>$ = 10,120~K and $<$log g$>$ = 3.547.  These average
values should better represent the mean atmospheric parameters during
the HB phase.  The dependence of the surface gravity
at \teff\ = 10,120~K on $Y$ is given in the top panel of Fig.~\ref{fig:L_g_Y}
for stars at the ZAHB, at the midpoint of the HB phase and at the point
90\% through the HB phase.  We see that the bulk of the HB phase
is spent within a narrow range in log g at a given helium abundance.
Only near the end of the HB phase does the surface gravity decrease
significantly below its ZAHB value.  The corresponding variation in the
luminosity with $Y$ is shown in the lower panel of Fig.~\ref{fig:L_g_Y}.  We 
again see that most of the HB phase is spent close to the ZAHB luminosity
as long as $Y$ $\lesssim$ 0.35.  At higher helium abundances an HB star
will evolve along a blue loop that becomes increasingly more extended
in effective temperature and that deviates more and more from the
ZAHB.  This leads to the substantial increase in the predicted
luminosity width of the HB evident in Fig.~\ref{fig:L_g_Y}.

\begin{figure}[h]
\includegraphics[width=\columnwidth,angle=0]{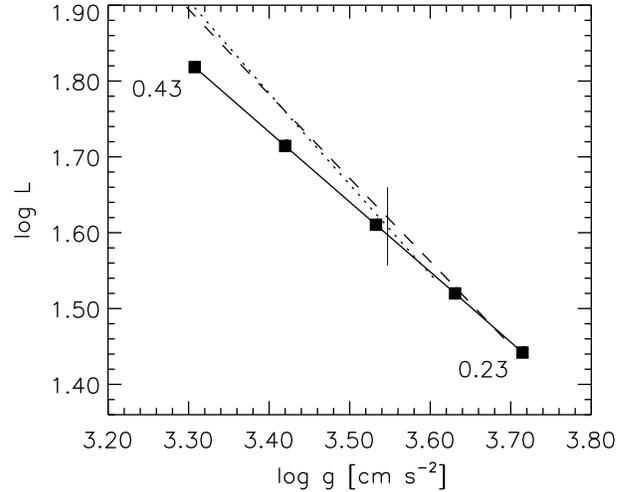}
\caption[]{Variation of the luminosity of HB stars at \teff\ = 10,120 K
with surface gravity over a range in the helium abundance
from $Y$ = 0.23 to $Y$ = 0.43.  The solid curve refers to
the ZAHB, the dashed curve to the midpoint of the HB phase,
and the dotted curve to the point 90\% through the HB
phase.  The values of $Y$, indicated by the solid squares
along the ZAHB curve, increase from $Y$ = 0.23 in the lower
right to $Y$ = 0.43 in the upper left in increments of 0.05.  The
thin vertical line at log g = 3.547 denotes the mean gravity
of the cool blue tail stars.
\label{fig:L_g}} 
\end{figure}

The results in Fig.~\ref{fig:L_g_Y} are combined in Fig.~\ref{fig:L_g}
to show the predicted variation of the luminosity with surface gravity
at \teff\ = 10,120~K.  The thin vertical line represents the mean
surface gravity of our cool blue tail stars.  We see that the
luminosity at this surface gravity does not depend very much on
whether the cool blue tail stars are near the ZAHB or near the end of
the HB phase.  Assuming that, on average, they are near the midpoint
of the HB phase, we find a mean luminosity $<$log L$>$ of 1.62 for
\logg\ = 3.547.  The
corresponding luminosity of a canonical HB star at the midpoint of its
HB phase is log L = 1.46 at \teff\ = 10,120~K.  The difference in
these luminosities implies that our cool blue tail stars are $\approx$0\fm4
brighter than canonical HB stars.  This result is in reasonable
agreement with the 0\fm5 increase in luminosity implied by the
upward slope of the HB in NGC~6388.  While our sample of cool blue
tail stars is admittedly very small and the errors in their surface
gravities significant, our results do suggest a brighter than
canonical luminosity that is at least consistent with the scenarios
that can explain the HB slope in NGC~6388.

One can also ask what helium abundance is required to produce the
lower surface gravities of the cool blue tail stars.  From the top
panel of Fig.~\ref{fig:L_g_Y} we find that a mean surface gravity of
log g = 3.547 corresponds to a helium abundance of $Y$ = 0.32 if the
cool blue tail stars are in the main part of their HB phase.  If,
however, they are near the end of their HB phase, then the required
helium abundance would be $Y$ = 0.28.  In either case a helium
enrichment, as might be expected from the helium pollution scenario,
could account for the properties of the cool blue tail stars in our
sample. Accounting for such an increased helium abundance in the
theoretical model spectra, which were used to determine effective
temperatures and surface gravities, would lower the derived surface
gravities, because an increase in helium abundance broadens the Balmer
line profiles. Thus a given observed profile will yield a lower
surface gravity if fitted with helium-enriched model spectra.

\subsection{Blue Tail Stars Hotter Than 12,000~K \label{blue_tail_hot}}

The hot blue tail stars in Fig.~\ref{fig:tg} have surface gravities
that are substantially smaller than the canonical values and indeed
even smaller than the surface gravities along the helium-rich ZAHB for
$Y$ = 0.43. As in the case of the cool blue tail stars, we can
use the masses obtained from the atmospheric parameters to test the
reliability of these surface gravities. The masses of the hot blue
tail stars plotted in Fig.~\ref{fig:tm} are, on average, about a
factor of 2 smaller than the theoretically predicted masses and,
moreover, are strongly correlated with the offset in log g from the
ZAHB in Fig.~\ref{fig:tg}. The only exception is star 2897 at \teff\ =
18,700~K, which has a mass in good agreement with the theoretical
value and which lies close to the $Y$ = 0.43 ZAHB in
Fig.~\ref{fig:tg}.  The atmospheric parameters of star 2897, if
representative of the true atmospheric parameters of a typical hot
blue tail star, would be consistent with an increase in the helium
abundance along the blue tail. However, the surface gravities of the
other hot blue tail stars seem spuriously too low.

The fact that the cool blue tail stars, which are visually
about 2 magnitudes brighter than the hot blue tail stars, have
reasonable parameters in Figs. 5 and 6 suggests that
the problems with the parameters of the hot blue tail stars
may be due to problems with background subtraction.  A possible
effect would be a filling of the Balmer line cores by
residual light from cool stars, which would simulate a higher
effective temperature (= less deep cores) for the hotter stars.
The wings would be less affected by such light, and the surface
gravity determined from them would probably be closer to
the true value.  However, with our current data we have no
way of verifying this possibility.

For comparison we show also results for stars in the metal-poor
globular cluster NGC~6752, which were analysed in the same way as the
stars discussed here, but are almost uncrowded. Obviously the stars in
NGC~6388 show lower surface gravities at a given temperature than
those in NGC~6752, whereas canonical stellar evolutionary theory would
argue in the opposite direction. This may be taken as support for the
influence of background subtraction. On the other hand, however, one
should keep in mind that the helium enrichment required to produce hot
HB stars in a metal-rich globular cluster would produce significantly
higher luminosities and thus lower gravities.

Blue tail stars hotter than 12,000~K are also affected by radiative
levitation which can increase the atmospheric metallicity to
super-solar values (Behr et al. \cite{beco99}, Moehler et
al. \cite{mosw00}, Moehler \cite{moeh01}, Behr \cite{behr03}).  We
attempted to include this effect in our analysis by using metal-rich
stellar atmospheres when determining the atmospheric parameters of the
hot blue tail stars.  However, these stellar atmospheres assume a
scaled solar abundance distribution which may not adequately account
for the highly non-solar abundance distribution produced by radiative
levitation. Further study is needed to see if this difference between
the assumed and actual abundance distributions can partly explain the
low gravities of the hot blue tail stars. This explanation is
supported by the low gravities found for hot blue tail stars in NGC~6752
(Moehler et al. \cite{mosw00}) and M~13 (Moehler et
al. \cite{mola03}).
It is important to note that the low helium abundances observed in the
hot blue tail stars are caused by diffusion effects and do not
contradict any helium pollution scenario.

\begin{figure}[h]
\includegraphics[width=\columnwidth,angle=0]{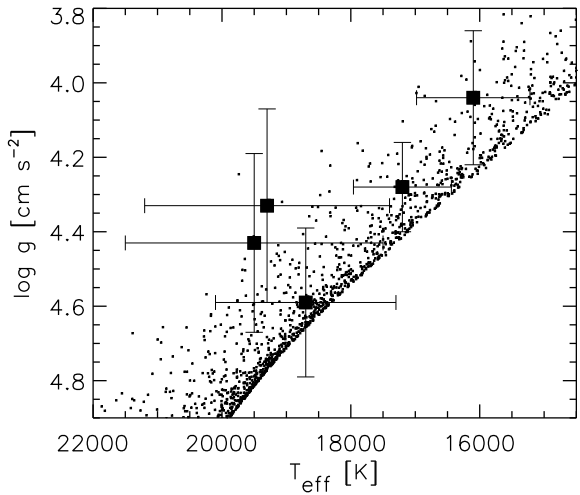}
\caption[]{Comparison of the atmospheric parameters
of the hot blue tail stars with an HB simulation for
a helium abundance $Y$ = 0.43.
\label{fig:HB_sim}} 
\end{figure}

Fig.~\ref{fig:tg} compares the surface gravities of the hot blue
tail stars only with the surface gravities of ZAHB models.  Since at high
helium abundances the HB evolutionary tracks can deviate considerably
from the ZAHB, we also need to examine how the post-ZAHB evolution
affects the predicted gravities.  In Fig.~\ref{fig:HB_sim} we overplot
the hot blue tail stars onto an HB simulation covering the entire HB
phase for $Y$ = 0.43.  For simplicity a uniform distribution in mass
was assumed in this simulation.  The hottest star in our sample (star
2612) is not included in Fig.~\ref{fig:HB_sim} because its surface
gravity is much too low to be explained by HB evolution.  Not
surprisingly, star 2897 lies within the bulk of the HB stars in
Fig.~\ref{fig:HB_sim}.  However, all of the other hot blue tail stars
lie in regions that are poorly populated.  We conclude that HB
evolution is unlikely to account for the low surface gravities of the
hot blue tail stars - a further indication of the uncertainty in the
atmospheric parameters of these stars.

\section{Conclusions}\label{sec:conclusions}

Our results for the cool blue tail stars in NGC~6388
are consistent with both the upward slope of the HB in
NGC~6388 and the predictions of the helium pollution
scenario.  Moreover, they resolve the puzzling conundrum
discussed by Moehler et al. (\cite{mosw99}), who found
higher than canonical surface gravities for a sample of
blue tail stars in NGC~6388 and NGC~6441 in contradiction
to all scenarios for explaining the upward sloping HBs
in these globular clusters.  Most likely, these large
gravities can be attributed to problems with the background
subtraction from the spectra.

Our results for the hot blue tail
stars remain problematical.  The low surface gravities
and masses found for these stars might be due to problems
with the background subtraction or possibly to inadequacies
in the stellar atmospheres used in the analysis.  Further
work is needed to clarify this point.  Unfortunately
the atmospheric parameters for the hot blue tail stars
do not permit a stringent test of the helium pollution
scenario.  In particular, the present data
cannot test the prediction for a greater gravity offset from
the canonical ZAHB in the hot blue tail stars compared to
the cool blue tail stars, as one would expect
if the helium abundance increased along the blue tail.

The present study illustrates the difficulties encountered when
doing spectroscopy in crowded fields.  From our experience we
conclude that the problem of the hot stars in NGC~6388 cannot be
solved by means of current ground-based observations.  One would need
either optical spectra at a {\em much} better spatial resolution
(around 0\farcs1) or multi-colour photometry including the near and
far UV.  Spectroscopy at such high spatial resolution is currently
available only at near-IR wavelengths, which would worsen the contrast
between the hot and cool stars and, in addition, would allow only the use of
hydrogen lines, for which the model atmospheres have not been
tested\footnote{IR observations of hot stars are usually done for
massive stars, not for evolved low-mass stars}. Multi-colour
photometry might provide an estimate of the effective temperatures
and luminosities of these stars, but as they are hot, the satellite UV
is definitely required.

\begin{acknowledgements}We gratefully acknowledge the efforts of the ESO
staff at Paranal and Garching that made these observations
possible. We also thank the referee, Dr. Vittoria Caloi, for a
  helpful report, that was delivered in very short time.
\end{acknowledgements}

\end{document}